\documentclass[11pt,a4paper]{article}
\pdfoutput=1

\usepackage{amsmath,amssymb,tikz}
\usepackage{graphicx}
 \allowdisplaybreaks
\def\be{\begin{equation}}
\def\ee{\end{equation}}
\def\ba#1\ea{\begin{align}#1\end{align}}
\def\bg#1\eg{\begin{gather}#1\end{gather}}
\def\bm#1\em{\begin{multline}#1\end{multline}}
\def\bmd#1\emd{\begin{multlined}#1\end{multlined}}

\def\({\left(}
\def\){\right)}
\def\[{\left[}
\def\]{\right]}

\def \be {\begin{equation}}
\def \ee {\end{equation}}
\def \ba {\begin{array}}
\def \ea {\end{array}}
\def \bea{\begin{eqnarray}}
\def \eea{\end{eqnarray}}

\def\bea{\begin{eqnarray}}
\def\eea{\end{eqnarray}}

\newcommand{\bit}{\begin{itemize}}  \newcommand{\eit}{\end{itemize}}
\newcommand{\ben}{\begin{enumerate}}  \newcommand{\een}{\end{enumerate}}

\long\def\symbolfootnote[#1]#2{\begingroup%
\def\thefootnote{\fnsymbol{footnote}}\footnote[#1]{#2}\endgroup}

\newcommand{\sysu}{{\it School of Physics and Astronomy, Sun Yat-Sen University, 2 Daxue Road, Zhuhai 519082, China}}

\begin{document}
\thispagestyle{empty}
\begin{center}

~\vspace{20pt}

{\Large\bf Anomalous Current Due to Weyl Anomaly for Conformal Field Theory}

\vspace{25pt}

Jiang-Jin Zheng, Dongqi Li, Yu-Qian Zeng,  Rong-Xin Miao ${}$\symbolfootnote[1]{Email:~\sf
  miaorx@mail.sysu.edu.cn}

\vspace{10pt}${}$\sysu

\vspace{2cm}

\begin{abstract}
Recently it is found that Weyl anomaly leads to new anomalous currents in an external electromagnetic field in a curved spacetime.  For simplicity, the initial works mainly focus on weak gravitational fields and the anomalous current is obtained for conformally flat spaces with small scale factors. In this paper, we generalize the results to the case with arbitrary scale factors. Firstly, we derive the transformation law of current under Weyl transformation, from which one can read off the anomalous current in general conformally flat spaces.  Secondly, by using the Weyl transformation of currents we provide a new derivation of the Weyl-anomaly-induced current near the boundary. Thus we have obtained the two kinds of anomalous currents in the literature from a unified formula. Finally, we extend the discussions to n-form fields and find similar anomalous currents.
\end{abstract}

\end{center}

\newpage
\setcounter{footnote}{0}
\setcounter{page}{1}

\tableofcontents

\section{Introduction}

The anomaly-induced current has drawn much attention in the past few years \cite{review}. The famous ones are chiral magnetic effect (CME) \cite{Vilenkin:1995um,
Vilenkin:1980fu, Giovannini:1997eg, alekseev, Fukushima:2012vr} and chiral vortical effect (CVE) \cite{Kharzeev:2007tn,Erdmenger:2008rm,
 Banerjee:2008th,Son:2009tf,Landsteiner:2011cp,Golkar:2012kb,Jensen:2012kj}, which refer
to the generation of currents due to an external magnetic field and the rotational motion in the charged fluid, respectively. Recently, it is found that Weyl anomaly can also produce anomalous current in an external electromagnetic field \cite{Chernodub:2016lbo, Chernodub:2017jcp,Chu:2018ksb,Chu:2018ntx}. See also \cite{Chernodub:2019blw,Chernodub:2018ihb,Chu:2018fpx,Chu:2019rod} for following works. It is remarkable that
a direct measurement of the beta function is proposed in  \cite{Chernodub:2019blw}.

Weyl anomaly measures the breaking of scaling symmetry of conformal field theory (CFT) due to quantum effects \cite{Duff:1993wm},
\begin{eqnarray}\label{definitionA}
\mathcal{A}= \partial_{\sigma} I_{\text{eff}}[e^{2\sigma} g_{ij}]|_{\sigma=0},
\end{eqnarray}
where $I_{\text{eff}}$ is the effective action of CFT and $g_{ij}$ are the metrics.  For 4d CFTs, Weyl anomaly takes the form \cite{Duff:1993wm}
\begin{eqnarray}\label{Weylanomaly}
\mathcal{A}=\int_M dx^4\sqrt{g} [c\ C^{ijkl}C_{ijkl}-a\ E_4+ b_1 F_{ij}F^{ij}]
\end{eqnarray}
where $C_{ijkl}$ are the Weyl tensors, $E_4=R_{ijkl} R^{ijkl} - 4 R_{ij} R^{ij} + R^2$ is the Euler density, $F=dA$ is the field strength and $c,a,b_1$ are central charges.

There are two kinds of anomalous current induced by Weyl anomaly in an external electromagnetic field \cite{Chernodub:2016lbo, Chernodub:2017jcp,Chu:2018ksb,Chu:2018ntx}. The first kind occurs in a curved spacetime without boundaries \cite{Chernodub:2016lbo, Chernodub:2017jcp}, while the second kind occurs in the general spacetime (including flat space) with boundaries \cite{Chu:2018ksb,Chu:2018ntx}. Below we call them Type I anomalous current and Type II anomalous current, respectively.

For a CFT defined in a conformally flat space
\begin{eqnarray}\label{conformallyflat}
ds^2= e^{2\sigma(x)}\delta_{ij}dx^i dx^j,
\end{eqnarray}
Type I anomalous current takes the form \cite{Chernodub:2016lbo, Chernodub:2017jcp}
\begin{eqnarray}\label{typeIcurrent}
J^i=4b_1 F^{ij}\partial_j \sigma +O(\sigma^2),
\end{eqnarray}
if the classical current vanishes $\nabla_i F^{ij}=0$.
Note that \cite{Chernodub:2016lbo, Chernodub:2017jcp} have assumed that the current  vanishes in a flat space (at least in some region), and they derive only the anomalous current in a weak gravitational background with small scale factor $\sigma$. It should be mentioned that a formal formula of the Type I anomalous current is obtained for general background fields in  \cite{Chernodub:2017jcp}. However, since the Green function is unknown generally, it is not clear how to derive the exact result of current from the formal formula of  \cite{Chernodub:2017jcp}.

For a CFT defined in a space with a boundary (BCFT) \footnote{In fact every spacetime has a conformal boundary. In this sense, all CFTs are also BCFTs. However,  the conformal boundary is a boundary located at infinity. While in this paper, by BCFT we mean the CFT defined in a manifold with the boundary which is located at a finite place. Note that BCFTs have fewer symmetries than CFTs due to the obstruction of boundaries. },
Type II anomalous current takes the universal form \cite{Chu:2018ksb,Chu:2018ntx}
\begin{eqnarray}\label{typeIIcurrent}
J^i=\frac{4b_1 F^{ij}n_j}{x} +..., \ x\sim 0,
\end{eqnarray}
 where $x$ is the proper distance to the boundary, $n_i$ are the normal vectors and $...$ denote higher order terms in $O(x)$. Note that there are boundary contributions to the current density, which can exactly cancel the  apparent ``divergence'' in the bulk current (\ref{typeIIcurrent}) at $x=0$ and define a finite total current \cite{Chu:2018ksb}. Note also that (\ref{typeIIcurrent}) actually applies to the general boundary quantum field theory instead of only BCFT \cite{Chu:2018ksb}. For simplicity, we focus on CFT/BCFT in this paper.

 In this paper, we aim to obtain the Type I anomalous current for general conformally flat spaces and try to reveal relations between these two kinds of anomalous currents. We find that the Weyl transformations of currents play an important role. By applying the field-theoretical and holographic methods respectively, we find that the current of CFTs transform as
 \begin{eqnarray}\label{holcurrentbar1Intr}
J'_i=e^{-2\sigma} J_i+ 4 b_1 \nabla'_j( F'{}_i^{\ j} \sigma),
\end{eqnarray}
 under the Weyl transformations
 \begin{eqnarray}\label{Weyl transformation}
g'_{ij} = e^{2\sigma} g_{ij}.
\end{eqnarray}
Here $\nabla'$ and $F'$ are the covariant derivative and the field strength defined with respect to the metric $g'_{ij}$, respectively.
From (\ref{holcurrentbar1Intr}) and the assumption that $J_i=0$ in some region of flat spaces \cite{Chernodub:2016lbo, Chernodub:2017jcp}, we obtain Type I anomalous current
\begin{eqnarray}\label{typeIcurrentIntr}
J'{}^i_{\text{anomaly}}=4b_1 \nabla'_j(F'^{ij} \sigma)
\end{eqnarray}
in the same region \footnote{Suppose that $J_i=0$ in the region $f(x^i)\le 1$ in a flat space. By ``the same region'' in conformally flat space, we means the same coordinate region $f(x^i)\le 1$. }
of conformally flat spaces with arbitrary scale factors. It is remarkable that BCFT in a half space is conformally equivalent to CFT in the Poincare patch of AdS. As a result, we can derive the Type II anomalous current (\ref{typeIIcurrent}) in a half space from the Type I anomalous current (\ref{typeIcurrentIntr}) with $\sigma=\ln x$.  Please see section 3 for details. This shows that there are close relations between these two kinds of anomalous currents, which is unified by the transformation law of currents (\ref{holcurrentbar1Intr}). Finally, we generalize our results to n-form fields in $2(n+1)$ dimensions and find similar anomalous currents
\begin{eqnarray}\label{anomalous currentnform Intro}
&&\text{ Type I:}\ J'_{\text{anomaly}\ i_1...i_n}=-2(n+1) b_n \nabla'_j(H'{}^{j}_{\ i_1...i_{n}}\sigma).\\
&&\text{ Type II:}\ J_{ i_1...i_n}= -2(n+1) b_n \frac{H{}^{x}_{\ i_1...i_{n}}}{x}+..., \ x\sim 0, \label{anomalous currentnformII Intro}
\end{eqnarray}
where $H=dB$ is the field strength of n-form field and $b_n$ are the central charges of Weyl anomaly (\ref{anomalynform}).

The paper is organized as follows.
In section 2, we derive the transformation law of current under Weyl transformation in AdS/CFT. From this transformation law, we obtain the anomalous current in arbitrary conformally flat space. In section 3, by using the conformal equivalence between AdS and flat space, we give a new derivation of Weyl anomaly induced current for BCFT.  In section 4, we extend our results to n-form fields. Finally, we conclude with some open questions in section 5.  For simplicity, we focus on Euclidean signature in this paper, the generalization to Lorentzian signature is straightforward.

\section{A derivation of Type I anomalous current}

In this section, we take field-theoretical and holographic methods to derive the transformation law of currents under Weyl transformations. By using this transformation law, we obtain the Type I anomalous current in conformally flat spaces with general scale factors. This is a generalization of the works of \cite{Chernodub:2016lbo, Chernodub:2017jcp}.

\subsection{Field-theoretical method}

Let us first discuss the field-theoretical method. Under Weyl transformation (\ref{Weyl transformation}), the effective action of CFTs transforms as
 \begin{eqnarray}\label{WeyleffI}
I'_{\text{eff}}(g'_{ij})=I_{\text{eff}}(g_{ij})+I_{\text{anomaly}}(g_{ij},\sigma),
\end{eqnarray}
where $I_{\text{anomaly}}$ is the anomalous action due to Weyl anomaly. From (\ref{WeyleffI}) we get the transformation law of current as
 \begin{eqnarray}\label{WeylcurrentIsec2}
J'^i=\frac{1}{\sqrt{g'}}\frac{\delta I'_{\text{eff}}}{\delta A_i}=e^{-4 \sigma} J^i+\frac{1}{\sqrt{g'}}\frac{\delta I_{\text{anomaly}}}{\delta A_i}.
\end{eqnarray}
Now the rest task is to derive anomalous action. For our purpose, we focus on the contributions relevant to the background field strength $F$. Only such terms can produce a current. From the definition of Weyl anomaly (\ref{definitionA}) and (\ref{Weylanomaly}), we have
 \begin{eqnarray}\label{dAnformsec2}
\delta_{\sigma} I_{\text{eff}}=I_{\text{anomaly}}(g_{ij},\delta\sigma)=\mathcal{A} \delta \sigma=b_1\int_M dx^4 \sqrt{g} F^{ij}F_{ij} \delta \sigma(x).
\end{eqnarray}
Note that $\sqrt{g}F^{ij}F_{ij}=\sqrt{g'} F'^{ij}F'_{ij}$ is independent of the scale factor. Here $F'_{ij}=F_{ij}=\partial_i A_{j}-\partial_j A_{i}$ and $F'^{ij}=F_{lk}g'^{li}g'^{kj}=e^{-4\sigma}F^{ij}$. Performing the integral in (\ref{dAnformsec2}), we get
 \begin{eqnarray}\label{Ieffsec2}
 I_{\text{anomaly}}=b_1\int_M dx^4 \sqrt{g'}  F'^{ij}F'_{ij} \sigma(x).
\end{eqnarray}
The above effective action can also be obtained by the method of dimensional regularization. Following \cite{Brown:1977sj,Herzog:2015ioa}, we have
 \begin{eqnarray}\label{Ieffsec2two}
I_{\text{anomaly}}&=&\lim_{\epsilon\to 0} \frac{b_1}{\epsilon} \int_M dx^{4+\epsilon}\left( \sqrt{g'}  F'^{ij}F'_{ij}- \sqrt{g}  F^{ij}F_{ij} \right)\nonumber\\
 &=& b_1\int_M dx^4 \sqrt{g'}  F'^{ij}F'_{ij} \sigma(x)
\end{eqnarray}
which reproduces (\ref{Ieffsec2}). Substituting the effective action (\ref{Ieffsec2}) into (\ref{WeylcurrentIsec2}), we finally get
 \begin{eqnarray}\label{currentsec2}
J'^{i}=e^{-4\sigma}J^{i}+ b_1 \nabla'_j(F'{}^{i j}\sigma).
\end{eqnarray}
Now we finish the derivations of the Weyl transformation law of current for general 4d CFTs. Please see the appendix for another derivation of the transformation law (\ref{currentsec2}), which shows that it agrees with the formal formula of the Type I anomalous current of \cite{Chernodub:2017jcp}.

Comments are in order. First, (\ref{currentsec2}) works for the general gravitational background and scale factors. In particular, the scale factor $\sigma$ need not be small. Second, the first term of RHS of (\ref{currentsec2}) is the ordinary transformation law, while the second term is due to Weyl anomaly and we call it the anomalous current
\begin{eqnarray}\label{anomalous current}
J'_{\text{anomaly}\ i}= 4 b_1 \nabla'_j(F'{}_i^{\ j}\sigma).
\end{eqnarray}
Note that the anomalous current is universal, in the sense that it depends on only the central charge of CFT instead of the states and temperature of the theory. Third, consider the case where $J_i =0$ in some region of a flat space, then the anomalous current contribute to all of the currents in the same region in general conformally flat space. This is a generalization of the works of \cite{Chernodub:2016lbo,Chernodub:2017jcp} to the general scale factor. Fourthly, (\ref{currentsec2}) implies that the currents obey the Wess-Zumino-like condition
\begin{eqnarray}\label{ Wess-Zumino}
[\delta_{\sigma_1},\delta_{\sigma_2}] J_i=0,
\end{eqnarray}
which is consistent with the fact that Weyl group is abelian. This can be regarded as a test of our result. Finally, we use holographic method to re-derive (\ref{currentsec2}) in the next section. This is another support to our results.

\subsection{Holographic method}

In this subsection, we take Penrose-Brown-Henneaux (PBH) transformation \cite{Imbimbo:1999bj,Schwimmer:2000cu} to study how currents transform under  Weyl transformations in AdS/CFT \cite{Maldacena:1997re}. From this transformation law, we can read off the Type I anomalous current in general conformally flat spaces. According to  \cite{Imbimbo:1999bj,Schwimmer:2000cu}, the Weyl transformations of the boundary metric can be understood as a certain subgroup of bulk diffeomorphisms in AdS/CFT. This is the so-called PBH transformation. For simplicity, the PBH transformation is worked out up to the linear order of scale factor in \cite{Imbimbo:1999bj,Schwimmer:2000cu}. For our purpose, we need to generalize the results of \cite{Imbimbo:1999bj,Schwimmer:2000cu} to non-perturbative scale factor.

Let us start with the general higher derivative gravity and Maxwell theory in the bulk, whose action is given by
\begin{eqnarray}\label{bulkaction}
I=\int_M dX^5\sqrt{G} [f(R, \nabla R,...)+b_1 \mathcal{F}_{\mu\nu}\mathcal{F}^{\mu\nu}],
\end{eqnarray}
where $X^{\mu}=(\rho, x^i)$ are the bulk coordinates, $b_1$ is the central charge of Weyl anomaly, $\mathcal{F}=d\mathcal{A}$ is the field strength, $R_{\mu\nu\alpha\sigma}$ are the curvatures and `...' denote higher derivatives of curvatures. For simplicity, we have ignored the indexes of curvatures. We assume that the asymptotically AdS are solutions to the above theory.  In Feferman-Graham gauge, the metric of asymptotically AdS takes the form
\begin{eqnarray}\label{bulkmetric}
ds^2=\frac{d\rho^2}{4\rho^2}+\frac{\hat{g}_{ij}(x,\rho)dx^idx^j}{\rho},
\end{eqnarray}
where $\hat{g}_{ij}=g_{ij}+\rho g_{(1)ij}+...$. Similarly, we can choose a gauge for the bulk gauge field
\begin{eqnarray}\label{bulkvector1}
&&\mathcal{A}_{\rho}=0, \\
&&\mathcal{A}_{i}=A_{i}+\rho [A_{(1)i}+ \bar{A}_{(1)i}\ln\rho]+... \label{bulkvector2}
\end{eqnarray}
Here $g_{ij}$ and $A_{i}$ are boundary metrics and vectors, respectively.  Note that $\bar{A}_{(1)i}$ can be derived from either the bulk Maxwell equations or the Weyl anomaly
\begin{eqnarray}\label{Weylanomalycurrentsec2}
\bar{A}_{(1)i}=-\frac{1}{4}\nabla_j F^{j}_{\ i}.
\end{eqnarray}
Please see the appendix for the details.
According to \cite{Gynther:2010ed,Papadimitriou:2005ii}, the holographic current of CFT is given by
\begin{eqnarray}\label{holcurrent}
J^i=\lim_{\rho\to 0}\frac{-4 b_1}{\sqrt{g}}\sqrt{G}\mathcal{F}^{\rho i}.
\end{eqnarray}
From (\ref{bulkmetric},\ref{bulkvector1},\ref{bulkvector2},\ref{holcurrent}), we get
\begin{eqnarray}\label{holcurrent1}
J_i=-8 b_1A_{(1)i},
\end{eqnarray}
where we have subtracted the finite and log divergent term proportional to $ \bar{A}_{(1)i}$ following the standard approach of holographic renormalization. Please refer to the appendix for the details of calculations.

We aim to derive the transformation law of current under the Weyl transformations (\ref{Weyl transformation}).
According to \cite{Imbimbo:1999bj} , the Weyl transformations can be realized by suitable bulk diffeomorphisms. Inspired by \cite{Imbimbo:1999bj}, we take the ansatz
\begin{eqnarray}\label{diffeomorphisms1}
&&\rho=\rho' e^{-2\sigma(x')} \left(1+\sum_{n=1}^{\infty} \rho'^n b_{(n)}(x') \right)\\
&&x^i=x'^i+\sum_{n=1}^{\infty} \rho'^n a^i_{(n)}(x') \label{diffeomorphisms2}
\end{eqnarray}
We require that the above diffeomorphisms leave the form of bulk metric (\ref{bulkmetric}) invariant, i.e.,
\begin{eqnarray}\label{diffeomorphisms1ri}
&&G'_{\rho \rho}=\frac{\partial X^{\mu}}{\partial \rho'} \frac{\partial X^{\nu}}{\partial \rho'} G_{\mu\nu}=\frac{1}{4\rho'^2},\\
&&G'_{\rho i}=\frac{\partial X^{\mu}}{\partial \rho'} \frac{\partial X^{\nu}}{\partial x'^i} G_{\mu\nu}=0 .\label{diffeomorphisms2ri}
\end{eqnarray}
 Substituting (\ref{diffeomorphisms1},\ref{diffeomorphisms2}) into (\ref{diffeomorphisms2ri}), we derive
 \begin{eqnarray}\label{diffeomorphisms3ri}
G'_{\rho i}=\frac{1}{\rho'}(-\frac{1}{2} \partial_i \sigma+a_{(1)}^j g'_{ij} )+O(\rho'^0)=0,
\end{eqnarray}
from which we get
 \begin{eqnarray}\label{ai}
a_{(1)}^i=\frac{1}{2} g'^{ij}\partial_j \sigma.
\end{eqnarray}
Note that $g'{}_{}^{ij}=e^{-2\sigma}g_{}^{ij}$ is nonperturbative in the scale factor.
Similarly, we can calculate $a_{(n)}^i$ and $b_{(n)}$ order by order from (\ref{diffeomorphisms1ri},\ref{diffeomorphisms2ri}). Since they are irrelevant to the calculations of holographic current, we do not discuss them here.

Now we are ready to derive the transformation law of current under Weyl transformation. Under the diffeomorphisms  (\ref{diffeomorphisms1},\ref{diffeomorphisms2}), the bulk gauge fields become
\begin{eqnarray}\label{dA1}
\mathcal{A}'_{\rho}(\rho', x')&=&\frac{\partial X^{\mu}}{\partial \rho'}  \mathcal{A}_{\mu}(\rho, x)\nonumber\\
&=&a_{(1)}^i A_{i}+O(\rho'),
\end{eqnarray}
and
\begin{eqnarray}\label{dA2}
&&\mathcal{A}'_{i}(\rho', x')=\frac{\partial X^{\mu}}{\partial x'^i}  \mathcal{A}_{\mu} (\rho, x)\nonumber\\
&=&(\delta^j_i+\rho'\partial_i a_{(1)}^j)\left(A_{j}(x)+\rho [A_{(1)j}(x)+\bar{A}_{(1)j}\ln\rho]\right)+O(\rho'^2)\nonumber\\
&=&(\delta^j_i+\rho'\partial_i a_{(1)}^j)\left(A_{j}(x')+\rho' a_{(1)}^k\partial_k A_{j}(x')+\rho' e^{-2\sigma} [A_{(1)j}(x')+\bar{A}_{(1)j}(x')\ln(\rho' e^{-2\sigma})]\right)+O(\rho'^2)\nonumber\\
&=& A_{i}(x')+\rho'\left(   e^{-2\sigma} A_{(1)i}(x') +\partial_i a_{(1)}^jA_{j}(x')+a_{(1)}^j\partial_j A_{i}(x')-2\sigma e^{-2\sigma}\bar{A}_{(1)i} \right)+O(\rho' \ln \rho',\rho'^2).\nonumber\\
\end{eqnarray}
Note that we have used the gauge $\mathcal{A}_{\rho}=0$ in the above derivations. However $\mathcal{A}'_{\rho}$ becomes non-zero after the coordinate transformations. As a result, we cannot use the formula like (\ref{holcurrent1}) to calculate the current $J'_i$.  Instead, we make use of the general formula like  (\ref{holcurrent})
\begin{eqnarray}\label{holcurrentbar}
J'^i=\lim_{\rho'\to 0}\frac{-4 b_1}{\sqrt{g'}}\sqrt{G'}\mathcal{F'}^{\rho i},
\end{eqnarray}
from which we get
\begin{eqnarray}\label{holcurrentbar}
J'_i=-8 b_1\left(A'_{(1)i}-\partial_i A'_{(0)\rho}\right),
\end{eqnarray}
where $A'_{(0)\rho}=\lim_{\rho'\to 0} A'_{\rho}$.
Another method to derive the holographic current (\ref{holcurrentbar}) is to perform a gauge transformation to make $\mathcal{A}'_{\rho}=0$, i.e.,
\begin{eqnarray}\label{gaugetransformation}
\mathcal{A}'_{\mu} \to \mathcal{A}'_{\mu} -\partial_{\mu} \left(A'_{(0)\rho} \rho'+O(\rho'^2)\right).
\end{eqnarray}
And then we can use the formula like (\ref{holcurrent1}) to get the holographic current (\ref{holcurrentbar}).
Substituting (\ref{Weylanomalycurrentsec2},\ref{holcurrent1},\ref{ai},\ref{dA1},\ref{dA2}) into (\ref{holcurrentbar}), we finally obtain
\begin{eqnarray}\label{holcurrentbar1}
J'_i=e^{-2\sigma} J_i+ 4 b_1 \nabla'_j (F'{}_i^{\ j} \sigma),
\end{eqnarray}
which agrees with the field-theoretical result (\ref{currentsec2}).

\section{A new derivation of Type II anomalous current}

By applying the transformation law of current (\ref{currentsec2}), we give a new derivation of the Type II anomalous current (\ref{typeIIcurrent}) for BCFT in this section. Note that the transformation law (\ref{currentsec2}) is derived for CFTs in section 2 and it works well only in the bulk region for BCFT. As on the boundary, it is expected that the boundary current transforms trivially under Weyl transformations. That is because there is no boundary contribution to Weyl anomaly from the background field strength. For simplicity, we focus on the bulk current of BCFT below. We make some comments on the boundary current and the finite total current at the end of this section.

A key observation is that BCFT in a flat space with a plane boundary
\begin{eqnarray}\label{BCFTflat}
ds^2=dx^2+dy_a^2,\ \  x\ge 0
\end{eqnarray}
is conformally equivalent to CFT in the Poincare patch of AdS
\begin{eqnarray}\label{CFTAdS}
ds^2=\frac{dx^2+dy_a^2}{x^2},\ \  x\ge 0.
\end{eqnarray}
where $x=0$ are the boundary of half space for BCFT and the conformal boundary of AdS for CFT, respectively. Note that these two kinds of boundaries are quite different, since the proper distance from an inside point to the boundary is finite in a half space, while is infinite in AdS.

Let us first discuss the current in an external electromagnetic field in AdS (\ref{CFTAdS}).  By definition of current, we have
\begin{eqnarray}\label{dIdA}
\delta I_{\text{ren}} =\int_M dx^4 \sqrt{g} J^i \delta A_i=\int_M dx^4 \frac{J^i}{x^4} \delta A_i,
\end{eqnarray}
where $ I_{\text{ren}}$ is the renormalized effective action, which is finite by renormalization. Since both the effective action and the vectors are finite, so do their variations. As a result, $\frac{J^i}{x^4}$ is finite from (\ref{dIdA}). Performing the Weyl transformation (\ref{Weyl transformation}) with $e^{2\sigma}=x^2$ and using the  transformation law (\ref{currentsec2}), we get the current in the half space (\ref{BCFTflat}),
\begin{eqnarray}\label{generalcurrentpriemBCFT}
J_{\text{BCFT}}^{i}=\frac{J^i}{x^4}+\frac{4b_1 F'^{ix}}{x}+4b_1 \nabla'_jF'^{ij}\ln x=\frac{4b_1 F'^{ix}}{x}+..., \ \ x>0
\end{eqnarray}
which exactly agrees with the anomalous current (\ref{typeIIcurrent}) near the boundary. In the above derivation we have used the fact that $\frac{J^i}{x^4}$ is finite as implied by (\ref{dIdA}). Now we finish the derivation of type II anomalous current (\ref{typeIIcurrent}). 

The above discussions can be easily generalized to curved spaces. Following the above approach and taking into account the conformal equivalence between the curved half space in Gauss normal coordinates
\begin{eqnarray}\label{BCFTflat1}
ds^2=dx^2+h_{ab}(x,y)dy^ady^b,\ \  x\ge 0,
\end{eqnarray}
and asymptotically AdS metrics
\begin{eqnarray}\label{CFTAdS1}
ds^2=\frac{dx^2+h_{ab}(x,y)dy^ady^b}{x^2},\ \  x\ge 0,
\end{eqnarray}
we can derive (\ref{generalcurrentpriemBCFT}) for the general curved case. Note that the anomalous current $\frac{4b_1 F'^{ix}}{x}$ is universal, while the finite part $\frac{J^i}{x^4}$ of (\ref{generalcurrentpriemBCFT}) depends on the states and temperature of the theory.

To end this section, let us calculate the boundary current and show that the total current is finite for BCFT \cite{Chu:2018ksb}. From the gauge invariance, one can derive the conservation laws of current \cite{Chu:2018ksb}
\begin{eqnarray}\label{conservationlawsbulk}
&&\text{ Bulk}: \ \ \ \ \ \ \nabla_i J^i=0 ,\\
&&\text{Boundary}: \ D_a j^a= - J_n, \label{conservationlawsbdy}
\end{eqnarray}
where $J_n$ is the normal component of the bulk current,  $J^a$ and $D_a$ are boundary currents and boundary covariant derivatives, respectively. From (\ref{generalcurrentpriemBCFT}), we can read off
\begin{eqnarray}\label{normalcurrentsec3}
J_n=J_x=-b_1 D_aF^{ax}\ln x.
\end{eqnarray}
Substituting (\ref{normalcurrentsec3}) into
the boundary conservation law (\ref{conservationlawsbdy}), we obtain the boundary
current
\begin{eqnarray}\label{bdycurrentsec3}
j_b=4b_1 F_{b x} \ln \epsilon,
\end{eqnarray}
where $\epsilon$ is the cut-off of the distance to the boundary. Now the total current density becomes
\begin{eqnarray}\label{currenttotalsec3}
J_b =\frac{4 b_1 F_{b x}}{x}+\delta(x-\epsilon)  4 b_1 F_{b x}\ln\epsilon+O(1),
\end{eqnarray}
which yields a finite total current after the integral along x.

\section{Anomalous current for n-form fields}

In this section, we discuss the anomalous current for n-form fields in $2(n+1)$-dimensions. The Weyl anomaly due to a background of  n-form field is given by
\begin{eqnarray}\label{anomalynform}
\mathcal{A}=\int_M dx^d \sqrt{g}\  b_n H^{i_1i_2...i_{n+1}}H_{i_1i_2...i_{n+1}},
\end{eqnarray}
where $d=2(n+1)$, $b_n$ is the central charge and $H=dB$ is the field strength of n-form $B_{i_1i_2...i_n}$.

\subsection{Field-theoretical method}

Following the method of sect. 2, we first give a field-theoretical derivation of the anomalous currents for n-form fields.
From the definition of Weyl anomaly (\ref{definitionA}) and (\ref{anomalynform}), we get the variation of the anomalous action  
 \begin{eqnarray}\label{dAnform}
\delta_{\sigma} I_{\text{anomaly}}=\mathcal{A} \delta \sigma=b_n\int_M dx^d \sqrt{g} H^{i_1i_2...i_{n+1}}H_{i_1i_2...i_{n+1}} \delta \sigma(x),
\end{eqnarray}
 where $\sqrt{g} H^{i_1i_2...i_{n+1}}H_{i_1i_2...i_{n+1}}=\sqrt{g'} H'^{i_1i_2...i_{n+1}}H'_{i_1i_2...i_{n+1}}$ is independent of the scale factor. Performing the integral in (\ref{dAnform}), we get
 \begin{eqnarray}\label{Ieffnformsec5}
 I_{\text{anomaly}}=b_n\int_M dx^d \sqrt{g'}  H'^{i_1i_2...i_{n+1}}H'_{i_1i_2...i_{n+1}} \sigma(x).
\end{eqnarray}
Similar to the case of section 2, the above effective action can also be obtained by the method of dimensional regularization. We have
 \begin{eqnarray}\label{Ieffnformsec5two}
  I_{\text{anomaly}}&=&\lim_{\epsilon\to 0} \frac{b_n}{\epsilon} \int_M dx^{d+\epsilon}\left( \sqrt{g'}  H'^{i_1i_2...i_{n+1}}H'_{i_1i_2...i_{n+1}}- \sqrt{g}  H^{i_1i_2...i_{n+1}}H_{i_1i_2...i_{n+1}} \right)\nonumber\\
 &=& b_n\int_M dx^d \sqrt{g'}  H'^{i_1i_2...i_{n+1}}H'_{i_1i_2...i_{n+1}} \sigma(x)
\end{eqnarray}
which reproduces (\ref{Ieffnformsec5}).
 From (\ref{WeyleffI}) and the anomalous action (\ref{Ieffnformsec5}), we get
 \begin{eqnarray}\label{nformcurrent}
J'^{i_1...i_n}=\frac{1}{\sqrt{g'}}\frac{\delta I'_{\text{eff}}}{\delta B_{i_1...i_n}}=e^{-2(n+1)\sigma}J^{i_1...i_n}-2(n+1) b_n \nabla'_j(H'{}^{j i_1...i_{n}}\sigma),
\end{eqnarray}
where the first term comes from the usual Weyl transformation of current and the second term originates in Weyl anomaly. We name the second term of (\ref{holcurrentbar1nform})  as the anomalous current
\begin{eqnarray}\label{anomalous currentnform}
J'_{\text{anomaly}\ i_1...i_n}=-2(n+1) b_n \nabla'_j(H'{}^{j}_{\ i_1...i_{n}}\sigma).
\end{eqnarray}
Suppose $J_{i_1...i_n}=0$ in some region of a flat space, then all of the currents are given by the anomalous currents (\ref{anomalous currentnform}) in the same region of a conformally flat space (\ref{conformallyflat}).

Recall that BCFT in the flat half space (\ref{BCFTflat}) is conformally equivalent to CFT in the Poincare patch of AdS (\ref{CFTAdS}). Following approaches of section 3, from (\ref{holcurrentbar1nform}) we can derive Type II anomalous current near the boundary as
 \begin{eqnarray}\label{anomalous currentnformII}
J_{ i_1...i_n}= -2(n+1) b_n \frac{H{}^{x}_{\ i_1...i_{n}}}{x}+..., \ x\sim 0,
\end{eqnarray}
where $x$ is the distance to the boundary and $...$ denotes higher orders of $O(x)$. Note that (\ref{anomalous currentnformII}) agrees with the result of \cite{Chu:2018fpx,Chu:2019rod} for two-form fields. This can be regarded as a check of our results.

\subsection{Holographic method}

Following section 2, we give a holographic derivation of the anomalous currents for n-form fields in this subsection. Similar to (\ref{bulkaction}), we take the following bulk action
\begin{eqnarray}\label{bulkactionnform}
I=\int_M dX^{d+1}\sqrt{G} [f(R, \nabla R,...)+b_n \mathcal{H}_{\mu_1...\mu_{n+1}}\mathcal{H}^{\mu_1...\mu_{n+1}}],
\end{eqnarray}
which yields the expected Weyl anomaly (\ref{anomalynform}). Please refer to the appendix of \cite{Chu:2018fpx} for the calculations of holographic Weyl anomaly for 2-form fields. The generalizations to n-form fields are straightforward.  From action (\ref{bulkactionnform}), we can derive the holographic current as
\begin{eqnarray}\label{holcurrentnform}
J^{i_1...i_n}=\frac{1}{\sqrt{g}}\frac{\delta I}{\delta B_{i_1...i_n}}=\lim_{\rho\to 0}\frac{-2(n+1) b_n}{\sqrt{g}}\sqrt{G}\mathcal{H}^{\rho i_1...i_{n}},
\end{eqnarray}
where $\mathcal{H}=d\mathcal{B}$ is the bulk field strength. For simplicity, we focus on the following gauge
\begin{eqnarray}\label{bulkvector1nform}
&&\mathcal{B}_{\rho i_1...i_{n-1}}=0, \\
&&\mathcal{B}_{i_1...i_n}=B_{i_1...i_n}+\rho [B_{(1)i_1...i_n}+\bar{B}_{(1)i_1...i_n}\ln \rho]+... \label{bulkvector2nform}
\end{eqnarray}
Similar to the case of one form, we have
\begin{eqnarray}\label{barBsection5}
\bar{B}_{(1)i_1...i_n}=-\frac{1}{4}\nabla^j H_{ji_1...i_n}=-\frac{1}{4}e^{2\sigma}\nabla'^j H'_{ji_1...i_n}.
\end{eqnarray}
After suitable holographic renormalization, the holographic current (\ref{holcurrentnform}) becomes
\begin{eqnarray}\label{holcurrentnform1}
J_{i_1...i_n}=-4(n+1) b_n B_{(1)i_1...i_{n}}.
\end{eqnarray}

Performing the diffeomorphisms (\ref{diffeomorphisms1},\ref{diffeomorphisms2}), we have
\begin{eqnarray}\label{dB1}
\mathcal{B}'_{\rho i_1...i_{n-1}}(\rho', x')(\rho', x')&=&\frac{\partial X^{\mu}}{\partial \rho'}\frac{\partial X^{\mu_1}}{\partial x'_{i_1}}...\frac{\partial X^{\mu_{n-1}}}{\partial x'_{i_{n-1}}}   \mathcal{B}_{\mu \mu_1...\mu_{n-1}}(\rho, x)\nonumber\\
&=&a_{(1)}^i B_{i i_1...i_{n-1}}(x')+O(\rho'),
\end{eqnarray}
and
\begin{eqnarray}\label{dB2}
&&\mathcal{B}'_{i_1...i_{n}}(\rho', x')=\frac{\partial X^{\mu_1}}{\partial x'_{i_1}}...\frac{\partial X^{\mu_{n}}}{\partial x'_{i_{n}}}   \mathcal{B}_{\mu_1...\mu_{n}} (\rho, x)\nonumber\\
&=&(\delta^{j_1}_{i_1}+\rho'\partial_{i_1} a_{(1)}^{j_1})...(\delta^{j_n}_{i_n}+\rho'\partial_{i_n} a_{(1)}^{j_n})\left(B_{j_1...j_n}(x)+\rho [B_{(1)j_1...j_n}(x)+\bar{B}_{(1)j_1...j_n}(x)\ln \rho]\right)+O(\rho'^2)\nonumber\\
&=&(\delta^{j_1}_{i_1}+\rho'\partial_{i_1} a_{(1)}^{j_1})...(\delta^{j_n}_{i_n}+\rho'\partial_{i_n} a_{(1)}^{j_n})\left(B_{j_1...j_n}(x')+\rho' a_{(1)}^k\partial_kB_{j_1...j_n}(x')\right)+O(\rho'^2)\nonumber\\
&&+(\delta^{j_1}_{i_1}+\rho'\partial_{i_1} a_{(1)}^{j_1})...(\delta^{j_n}_{i_n}+\rho'\partial_{i_n} a_{(1)}^{j_n})\left(\rho' e^{-2\sigma} B_{(1)j_1...j_n}(x')+\rho' e^{-2\sigma} \bar{B}_{(1)j_1...j_n}(x')\ln(\rho'e^{-2\sigma})\right)\nonumber\\
&=& B_{i_1...i_n}(x')+\rho'\left(  \sum_{p=1}^n\partial_{i_p} a_{(1)}^{j_p}B_{i_1...j_p...i_n}(x')+a_{(1)}^j\partial_j B_{i_1...i_n}(x')\right)\nonumber\\
&&+ \rho'\left(   e^{-2\sigma} B_{(1)i_1...i_n}(x')-2\sigma e^{-2\sigma} \bar{B}_{(1)i_1...i_n}(x')\right)+O(\rho'\ln\rho',\rho'^2)
\end{eqnarray}
Note that $\mathcal{B}'_{\rho i_1...i_{n-1}}$ become non-zero after the diffeomorphisms. To preserve the gauge $\mathcal{B}'_{\rho i_1...i_{n-1}}=0$ (\ref{bulkvector1nform}), let us perform a gauge transformation
\begin{eqnarray}\label{gaugetransformationnform}
\mathcal{B}'_{\mu_1...\mu_{n+1}} \to \mathcal{B}'_{\mu_1...\mu_{n+1}} -n \partial_{[\mu_1} \left(a_{(1)}^i B_{|i |\mu_2...\mu_{n}]} \rho'+O(\rho'^2)\right),
\end{eqnarray}
which yields $\mathcal{B}'_{\rho i_1...i_{n-1}}=0$ and
\begin{eqnarray}\label{dB2new}
\mathcal{B}'_{i_1...i_{n}}(\rho', x')&=& B_{i_1...i_n}(x')+\rho'\left(   e^{-2\sigma} [B_{(1)i_1...i_n}-2\sigma \bar{B}_{(1)i_1...i_n}](x')+a_{(1)}^{i_1}H_{i_1...i_{n+1}}(x')\right)\nonumber\\
&&+O(\rho'\ln\rho',\rho'^2).
\end{eqnarray}
Substituting  (\ref{ai},\ref{barBsection5},\ref{dB2new}) into the formula  (\ref{holcurrentnform1}),  we finally get the transformation law of current under Weyl transformations for n-form fields
\begin{eqnarray}\label{holcurrentbar1nform}
J'_{i_1...i_n}=e^{-2\sigma} J_{i_1...i_n}- 2(n+1) b_n \nabla'_j(H'{}^{j}_{\ i_1...i_{n}}\sigma).
\end{eqnarray}
Following the same approach of sect. 4.1, we can obtain the Type I anomalous current (\ref{anomalous currentnform}) and Type II anomalous current (\ref{anomalous currentnformII}) for n-from fields.

\section{Conclusions and Discussions}

In this paper, we investigate the anomalous current due to Weyl anomaly for CFTs. In particular, we obtain the Type I anomalous currents in general conformally flat spaces and unify the two kinds of anomalous currents into one formula, i.e., the Weyl transformation law of current. By using the field-theoretical and holographic methods, we derive the Weyl-transformation law of currents, from which we can read off the Type I anomalous current in a general conformally flat space. This is a generalization of the works of \cite{Chernodub:2016lbo, Chernodub:2017jcp}, which mainly discuss the case of weak gravity with small scale factors. Furthermore, by applying the transformation law of current, we give a new derivation of the anomalous current near the boundary  \cite{Chu:2018ksb,Chu:2018ntx}, and reveal that there are close relations between the Type I anomalous current \cite{Chernodub:2016lbo, Chernodub:2017jcp} and Type II anomalous current \cite{Chu:2018ksb,Chu:2018ntx}. Finally, we extend our discussions to n-form fields and find similar anomalous currents. We notice that our results agree with Type II anomalous current for the two-form field \cite{Chu:2018fpx,Chu:2019rod}. For simplicity, in this paper we discuss only the current induced by an external electromagnetic field in four dimensions. It is interesting to study the anomalous current in higher dimensions. Besides, it is also interesting to search for the applications of our results to cosmology and condensed matter. Since our results work for arbitrary conformally flat spaces, now we can study the current for general Robertson-Walker metrics in cosmology. We hope we could address these problems in the future.

\section*{Acknowledgements}
We would like to thank Stefan Theisen and Chen-Te Ma
for useful discussions and comments. This work is supported by the funding of Sun Yat-Sen University.

\appendix

\section{Another derivation of the anomalous current}

In this appendix, we give another derivation of the Type I anomalous current (\ref{anomalous current}) and show that it agrees with the formal formula of \cite{Chernodub:2017jcp}.

According to \cite{Riegert:1984kt,Mazur:2001aa,Mottola:2006ew,Armillis:2009pq}, the effective anomalous action generated by one-loop quantum corrections takes the form
\begin{eqnarray}\label{anomalous action}
I_{\text{eff}}& = & \frac{1}{8} \int d^4 x \sqrt{g(x)}  \int d^4 y \sqrt{g(y)}  \\
& & \cdot H(x) G_4(x,y)  \left[ 2 c C^2(y) - a H(y) + 2b_1 F_{ij}(y) F^{ij}(y)\right],
\nonumber
\end{eqnarray}
where $C^2=C_{ijkl} C^{ijkl} $ is the squared Weyl tensor, $H=E_4-\frac{2}{3}\Box R$, $E_4=R_{ijkl} R^{ijkl} - 4 R_{ij} R^{ij} + R^2$ is the Euler density and $G_4(x,y) $ is the Green function of the differential operator
\begin{eqnarray}\label{greenfunction}
\Delta_4 = \nabla_i \left( \nabla^i \nabla^j + 2 R^{ij} - \frac{2}{3} R g^{ij}\right) \nabla_j.
\nonumber
\end{eqnarray}
By definition, we have
\begin{eqnarray}\label{greenfunction1}
\Delta_4(x) G_4(x,y)=\frac{\delta^{(4)}(x-y)}{\sqrt{g}}.
\end{eqnarray}

From the anomalous action (\ref{anomalous action}), one can derive a formal expression of the anomalous current as \cite{Chernodub:2017jcp}
\begin{eqnarray}\label{generalcurrent}
J^{i}(x) & = & \frac{1}{\sqrt{g(x)}}\frac{\delta I_{\text{eff}}}{\delta A_i(x)}\nonumber \\
& = & \frac{b_1}{\sqrt{g(x)}} \frac{\partial }{\partial x^j } \big[\sqrt{g(x)}  \, F^{ij} (x) \\
& & \cdot \int d^4 y \sqrt{g(y)} G_4(x,y) \left(E(y) - \frac{2}{3} \Box R(y)\right)\big],\nonumber
\end{eqnarray}
which applies to general gravitational background. However, since the exact expression of green function $G_4$ is unknown, one cannot use (\ref{generalcurrent}) directly to derive the anomalous current. Fortunately, we do not need the exact expression of Green function $G_4$ to derive the transformation law of current (\ref{currentsec2}).  Note that $\Delta_4$ is a conformal differential operator, which transforms
\begin{eqnarray}\label{Delta4transformation}
\Delta_4\to \Delta'_4=e^{-4\sigma}\Delta_4
\end{eqnarray}
under the Weyl transformation $g_{ij}\to g'_{ij}=e^{2\sigma}g_{ij}$. From (\ref{greenfunction1}) and (\ref{Delta4transformation}), it is not difficult to observe that the Green function is invariant under Weyl transformation, i.e.,
\begin{eqnarray}\label{Greentransformation}
G_4\to G'_4=G_4.
\end{eqnarray}
After some calculations, we obtain the Weyl transformation of $(E-\frac{2}{3}\Box R)$ as
\begin{eqnarray}\label{Htransformation}
(E-\frac{2}{3}\Box R) \to (E'-\frac{2}{3}\Box' R')=e^{-4\sigma} (E-\frac{2}{3}\Box R+4\Delta_4 \sigma).
\end{eqnarray}
The above two equations yield
\begin{eqnarray}\label{keyformula}
&&\int d^4 y \sqrt{g'(y)} G'_4(x,y) \left(E'(y) - \frac{2}{3} \Box' R'(y)\right)\nonumber\\
&=&4\sigma(x)+\int d^4 y \sqrt{ g(y)} G_4(x,y) \left(E(y) - \frac{2}{3} \Box R(y)\right).
\end{eqnarray}

Now we are ready to derive the transformation law of current. By using (\ref{generalcurrent},\ref{keyformula}) together with $F'^{ij}=e^{-4\sigma} F^{ij}$ and $\sqrt{g'}=e^{4\sigma}\sqrt{g}$, we finally obtain
\begin{eqnarray}\label{generalcurrentpriem}
J'^{i}=e^{-4\sigma} J^i+4b_1 \nabla'_j (F'^{ij} \sigma),
\end{eqnarray}
which exactly agrees with (\ref{currentsec2}). From the above equation, we can read off the Type I anomalous current (\ref{anomalous current}). Now we have shown that our results agree with the formal expression of Type I anomalous current (\ref{generalcurrent}) obtained in \cite{Chernodub:2017jcp}.


\section{Holographic renormalization of current}

In this appendix, we derive eqs.(\ref{Weylanomalycurrentsec2},\ref{holcurrent1}) for the holographic current in section 2.
The current of CFT is defined by
 \begin{eqnarray}\label{CFTcurrentsec2app}
J^i=\frac{1}{\sqrt{g}}\frac{\delta I_{eff}}{\delta A_i}
\end{eqnarray}
where $I_{eff}$ is the non-renormalized effective action of CFTs, which takes the form \cite{Chu:2018ksb}
\begin{eqnarray}\label{Ieffnornapp}
I_{eff}=..+\mathcal{A} \ln\frac{1}{\epsilon}+I_{ren},
\end{eqnarray}
where $...$ denote divergent terms, $\mathcal{A}$ is the Weyl anomaly, $\epsilon$ is the cutoff and $I_{ren}$ is the renormalized finite effective action. From (\ref{CFTcurrentsec2app}), we get the non-renormalized current
 \begin{eqnarray}\label{CFTcurrentsec2appnon}
J^i=...+\frac{1}{\sqrt{g}}\frac{\delta \mathcal{A}}{\delta A_i}\ln\frac{1}{\epsilon}+J^i_{ren}.
\end{eqnarray}
One need performing suitable renormalization in order to obtain the finite current $J^i_{ren}$. Below we focus on the holographic renormalization for 4d CFTs.

In AdS/CFT, the effective action of CFTs is given by the on-shell gravitational action (\ref{bulkaction}). From (\ref{CFTcurrentsec2app}), we get the holographic current (\ref{holcurrent})
 \begin{eqnarray}\label{holocurrentapp}
J^i=\lim_{\rho\to 0}\frac{1}{\sqrt{g}}\frac{\delta I}{\delta A_i}=\lim_{\rho\to 0}\frac{-4 b_1}{\sqrt{g}}\sqrt{G}\mathcal{F}^{\rho i}
\end{eqnarray}
Substituting the bulk vectors (\ref{bulkvector1},\ref{bulkvector2}) into (\ref{holocurrentapp}), we get
\begin{eqnarray}\label{holcurrent1app}
J_i=-8 b_1[A_{(1)i}+\bar{A}_{(1)i}+\bar{A}_{(1)i} \ln \epsilon^2],
\end{eqnarray}
where we have choosen the cut-off $\rho=\epsilon^2$. Comparing the log divergent terms in (\ref{CFTcurrentsec2appnon}) and (\ref{holcurrent1app}), we obtain
\begin{eqnarray}\label{relationapp}
16 b_1\bar{A}_{(1)}^{\ i}=\frac{1}{\sqrt{g}}\frac{\delta \mathcal{A}}{\delta A_i}=-4b_1 \nabla_j F^{ji},
\end{eqnarray}
where we have used (\ref{Weylanomaly}) in the last equation of (\ref{relationapp}). From (\ref{relationapp}), we obtain (\ref{Weylanomalycurrentsec2}) of section 2,
\begin{eqnarray}\label{Weylanomalycurrentsec2app}
\bar{A}_{(1)i}=-\frac{1}{4}\nabla_j F^{j}_{\ i}=-\frac{1}{4}e^{2\sigma}\nabla'_j F'^{j}_{\ i}.
\end{eqnarray}
Now we finish the derivation of $\bar{A}_{(1)i}$ from Weyl anomaly. Note that $\bar{A}_{(1)i}=e^{2\sigma} \bar{A'}_{(1)i}$ transforms trivially under Weyl transformations.

Following the standard approach of holographic renormalization, we can add suitable boundary counter terms to cancel the divergence of gravitational action and holographic  currents. Let us focus on the counter terms related to gauge fields. The renormalized gravitational action is given by \cite{Genolini:2016ecx}
\begin{eqnarray}\label{bulkactionrenapp}
I_{reg}&=&\int_M dX^5\sqrt{G} [f(R, \nabla R,...)+b_1 \mathcal{F}_{\mu\nu}\mathcal{F}^{\mu\nu}]\nonumber\\
&-&\mathcal{A}\ln \frac{1}{\epsilon}+
 \int_{\rho=\epsilon^2} dx^4\sqrt{\hat{g}} (\xi F_{ij}F_{kl}\hat{g}^{ik}\hat{g}^{jk}),\end{eqnarray}
where the first counter term is designed to cancel the log divergent term of the holographic current (\ref{holcurrent1app}), and the second counter term is finite, which depends on the choices of renormalization schemes labeled by the parameter $\xi$. From the renormalized gravitational action (\ref{bulkactionrenapp}) and (\ref{Weylanomalycurrentsec2app}), we derive the renormalized holographic current
\begin{eqnarray}\label{holcurrent1renapp}
J_i=-8 b_1A_{(1)i}+(16\xi-8b_1)\bar{A}_{(1)i},
\end{eqnarray}
which reduces to (\ref{holcurrent1}) of section 2 for $\xi=b_1/2$. One can easily check that the choices of $\xi$ do not affect the derivations of the transformation law of current (\ref{holcurrentbar1}) in section 2. That is because $\bar{A}_{(1)i}=e^{2\sigma} \bar{A'}_{(1)i}$ transforms trivially under Weyl transformations. Finally, it is straightforward to generalize the above discussions to n-form fields and derive eqs.(\ref{barBsection5},\ref{holcurrentnform1}) of section 5.

\end{document}